\documentclass[conference]{IEEEtran}
\IEEEoverridecommandlockouts
\usepackage{cite}
\usepackage{amsmath,amssymb,amsfonts}
\usepackage{algorithmic}
\usepackage{graphicx}
\usepackage{textcomp}
\usepackage{hyperref}
\usepackage{xcolor}
\usepackage{multirow}
\usepackage[linesnumbered,ruled,vlined]{algorithm2e}
\def\BibTeX{{\rm B\kern-.05em{\sc i\kern-.025em b}\kern-.08em
    T\kern-.1667em\lower.7ex\hbox{E}\kern-.125emX}}

\usepackage{listings}
\definecolor{codegreen}{rgb}{0,0.6,0}
\definecolor{codegray}{rgb}{0.5,0.5,0.5}
\definecolor{codepurple}{rgb}{0.58,0,0.82}
\definecolor{backcolour}{rgb}{0.99,0.99,0.99}
\usepackage{subcaption}
\definecolor{delim}{RGB}{20,105,176}
\definecolor{numb}{RGB}{106, 109, 32}
\definecolor{string}{rgb}{0.64,0.08,0.08}

\lstdefinestyle{qfaasstyle}{
    backgroundcolor=\color{backcolour},   
    commentstyle=\color{codegreen},
    keywordstyle=\color{magenta},
    numberstyle=\tiny\color{codegray},
    stringstyle=\color{codepurple},
    basicstyle=\ttfamily\footnotesize,
    breakatwhitespace=false,         
    breaklines=true,                 
    captionpos=b,
    title={},
    keepspaces=true,                 
    numbers=none,                    
    numbersep=10pt,                  
    showspaces=false,                
    showstringspaces=false,
    showtabs=false,                  
    tabsize=1,
    frame=single,
    xleftmargin=0.1in,
    xrightmargin=0.1in
}

\lstdefinelanguage{java}{
    rulecolor=\color{black},
    postbreak=\raisebox{0ex}[0ex][0ex]{\ensuremath{\color{gray}\hookrightarrow\space}},
    upquote=true,
    morestring=[b]",
    literate=
     *{0}{{{\color{numb}0}}}{1}
      {1}{{{\color{numb}1}}}{1}
      {2}{{{\color{numb}2}}}{1}
      {3}{{{\color{numb}3}}}{1}
      {4}{{{\color{numb}4}}}{1}
      {5}{{{\color{numb}5}}}{1}
      {6}{{{\color{numb}6}}}{1}
      {7}{{{\color{numb}7}}}{1}
      {8}{{{\color{numb}8}}}{1}
      {9}{{{\color{numb}9}}}{1}
      {\{}{{{\color{delim}{\{}}}}{1}
      {\}}{{{\color{delim}{\}}}}}{1}
      {[}{{{\color{delim}{[}}}}{1}
      {]}{{{\color{delim}{]}}}}{1},
}

\begin{document}

\title{DRLQ: A Deep Reinforcement Learning-based
Task Placement for Quantum Cloud Computing
}

\author{
    \IEEEauthorblockN{Hoa T. Nguyen\textsuperscript{1}, Muhammad Usman\textsuperscript{2,3}, and Rajkumar Buyya\textsuperscript{1}}
    \IEEEauthorblockA{\textsuperscript{1}\textit{Cloud Computing and Distributed Systems (CLOUDS) Laboratory}, \\ \textit{School of Computing and Information Systems, The University of Melbourne, Parkville, 3052, Victoria, Australia}}
    \IEEEauthorblockA{\textsuperscript{2}\textit{School of Physics, The University of Melbourne, Parkville, 3052, Victoria, Australia}}
    \IEEEauthorblockA{\textsuperscript{3}\textit{Data61, CSIRO, Clayton, 3168, Victoria, Australia}} 
    thanhhoan@student.unimelb.edu.au, \{muhammad.usman, rbuyya\}@unimelb.edu.au
    
}

\maketitle

\begin{abstract}
The quantum cloud computing paradigm presents unique challenges in task placement due to the dynamic and heterogeneous nature of quantum computation resources. Traditional heuristic approaches fall short in adapting to the rapidly evolving landscape of quantum computing. This paper proposes DRLQ, a novel Deep Reinforcement Learning (DRL)-based technique for task placement in quantum cloud computing environments, addressing the optimization of task completion time and quantum task scheduling efficiency. It leverages the Deep Q Network (DQN) architecture, enhanced with the Rainbow DQN approach, to create a dynamic task placement strategy. This approach is one of the first in the field of quantum cloud resource management, enabling adaptive learning and decision-making for quantum cloud environments and effectively optimizing task placement based on changing conditions and resource availability. We conduct extensive experiments using the QSimPy simulation toolkit to evaluate the performance of our method, demonstrating substantial improvements in task execution efficiency and a reduction in the need to reschedule quantum tasks. Our results show that utilizing the DRLQ approach for task placement can significantly reduce total quantum task completion time by 37.81\% to 72.93\% and prevent task rescheduling attempts compared to other heuristic approaches.  
\end{abstract}


\section{Introduction}
Quantum computing is at the forefront of technological innovation, with the potential to drive advances in fields such as cryptography \cite{pirandola2020advances}, finance \cite{egger2020quantum}, machine learning \cite{west2023towards}, and complex chemical simulation \cite{kandala2017hardware}. It has the capability to solve numerous problems that are currently intractable by classical computers. Besides, the emergence of quantum cloud computing \cite{nguyen2024quantum} represents a major advancement in providing access to the computational capabilities of quantum computing. This integration enables users worldwide to execute quantum algorithms on remotely accessible quantum computers, thereby overcoming the significant barriers of cost and physical access associated with quantum hardware \cite{kaiiali2019cloud}. The quantum cloud paradigm has not only extended the accessibility of quantum computing but has also presented new challenges and opportunities in optimizing the utilization of these cloud-based quantum computation resources.
\begin{figure}[!t]
\centering
\includegraphics[width=3.5in]{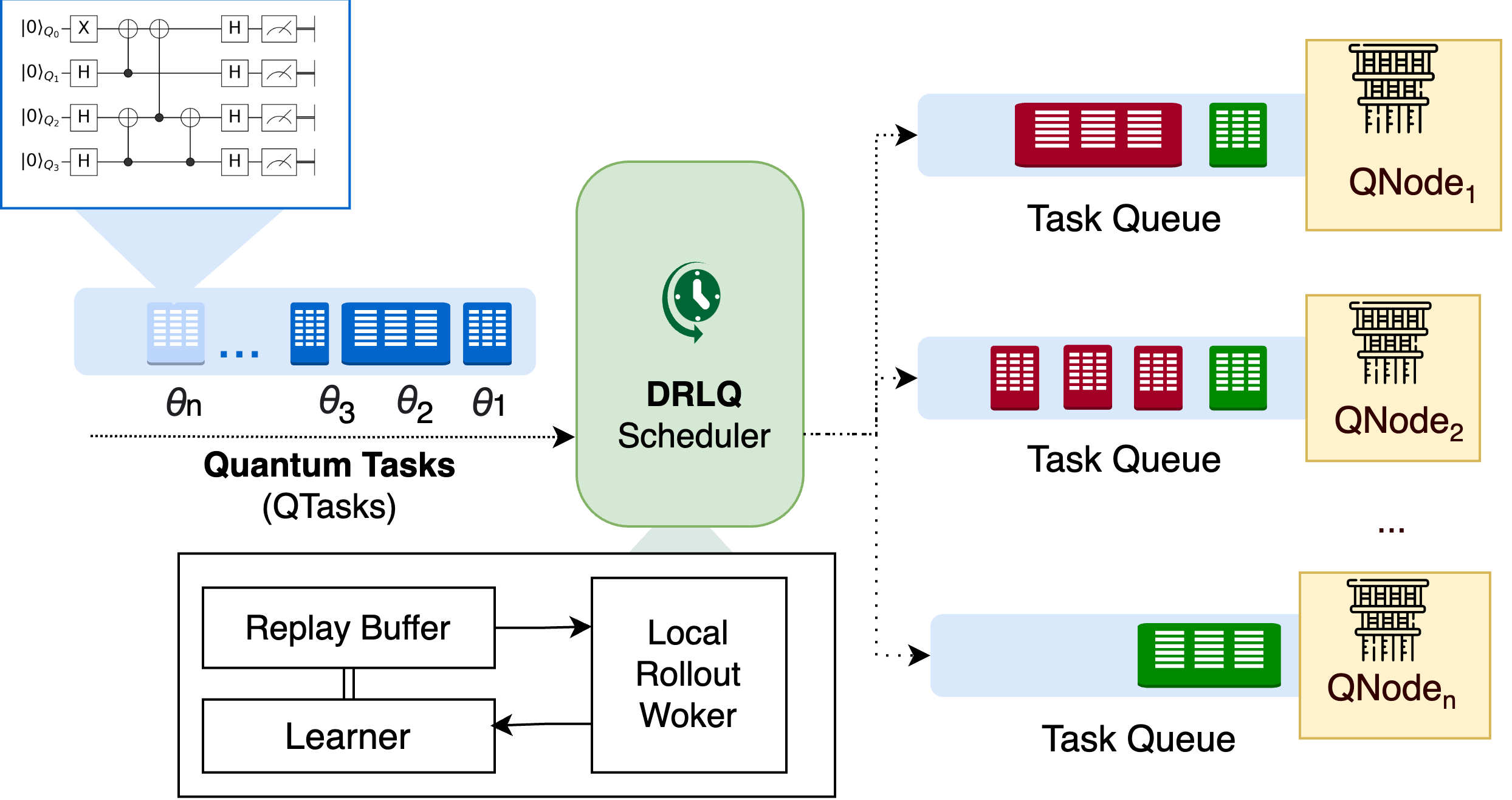}
\caption{Overview of the system model for the task placement problem in quantum cloud environments}
\label{fig_overview}
\end{figure}

Despite its significant potential, the efficient utilization of quantum cloud computing resources faces substantial challenges, particularly in the context of quantum cloud resource orchestration. Specifically, quantum task placement, i.e., selecting an appropriate quantum backend or physical hardware and associated parameters for executing quantum tasks, is crucial for the performance and reliability of quantum computations \cite{nguyen2024qfaas}. However, the current landscape of quantum task placement is denoted by dependence on heuristic approaches or manually crafted policies \cite{ravi_adaptive_2021}. Although practical in certain contexts, these approaches do not take full advantage of the dynamic capabilities of quantum cloud computing environments. They lack the flexibility and adaptability required to optimize performance in the face of ongoing advancements in quantum hardware and the increasing complexity of practical quantum applications. \cite{dalzell2023quantum}.

This scenario underscores the critical need for novel, adaptive techniques in resource management capable of harnessing the potential of quantum cloud computing. Integrating deep reinforcement learning (DRL) into the quantum task placement process presents a promising approach to address these challenges. By incorporating the principles of DRL, which is well-suited for navigating complex and dynamic environments such as cloud-edge in the classical domain with several successful examples such as \cite{hasselt2016deep, yi2020efficient, goudarzi2023distributed}, a DRL-based task placement strategy can potentially enhance quantum cloud resource management. As far as we know, this study is the first attempt to apply a deep reinforcement learning-based technique designed for the task placement problem in quantum cloud computing environments \cite{nguyen2024quantum}. Our approach aims to navigate the complexities of quantum systems dynamics, balancing performance metrics and adaptively learning the optimal task placement policy through continuous interactions with the quantum computing environment. Our proposed methodology leveraging Deep Q Networks (DQN) architecture and empowered by Rainbow DQN approach \cite{hessel2018rainbow}, which combines advantages of different DQN algorithms, including Double DQN, Prioritized Replay, Multi-step learning, Distributional RL, and Noisy Nets, seeks to optimize task placement in quantum cloud computing environment. 

The major contributions and novelty of our work are:
\begin{itemize}
    \item We propose one of the first applications of DRL techniques to address the task placement problem in quantum cloud computing, leveraging the enhanced combining improvement of the Rainbow DQN technique \cite{hessel2018rainbow} for robust and adaptive decision-making.
    \item Through extensive experimentation, we have shown that our DRLQ approaches can significantly reduce the total completion time by 37.81\% to 72.93\% and minimize the need for task rescheduling compared to other popular heuristic-based approaches.
    \item Our findings emphasize a possible method for tackling the problem of quantum task placement in order to optimize resource management in quantum cloud computing environments. This provides a starting point for additional thorough research that considers more quantum-specific properties, such as execution accuracy, quantum circuit transpilation, and quantum error rates.
\end{itemize}

The rest of the paper is organized as follows: Section II reviews related work, identifying gaps our research addresses. Section III describes the system model and problem formulation for task placement in quantum cloud computing, then explains our DRL model, focusing on the DQN architecture and Rainbow approach. Section IV evaluates our method's performance through simulations and discusses the implications of our findings. Section V concludes the paper by summarizing our contributions along with future work.

\section{Related Work}
The literature on task placement in quantum cloud computing is nascent, with only few works beginning to address the unique challenges posed by this emerging paradigm. This section briefly reviews the existing studies, highlighting the pioneering efforts in task placement within quantum cloud computing and the potential of DRL to innovate in this area. We summarise several related works in the quantum cloud domain and representative works in the classical cloud-edge domain in Table \ref{tab:related-work}.

\begin{table}[htbp]
\caption{Representative works related to our study}
\label{tab:related-work}
\begin{tabular}{|l|l|l|l|}
\hline
\textbf{References} & \textbf{\begin{tabular}[c]{@{}l@{}}Resource Management\\ Problem\end{tabular}} & \textbf{Environment} & \textbf{Approach} \\ \hline
\cite{chen2022adaptive} & Task Placement & Classical Cloud & DRL \\ \hline
\cite{goudarzi2023distributed} & Task Placement & Classical Edge & DRL \\ \hline
\cite{ravi_adaptive_2021} & Task Placement & Quantum Cloud & Heuristics \\ \hline
\cite{kaewpuang_stochastic_2022} & Qubit allocation & Quantum Cloud & Heuristics \\ \hline
\cite{cicconetti_resource_2022} & Resource Allocation & Quantum Network & Heuristics \\ \hline
\textit{\textbf{Our work}} & Task Placement & Quantum Cloud & DRL \\ \hline
\end{tabular}
\end{table}

The traditional task placement strategies used in classical cloud computing cannot be directly applied to the quantum context due to differences in quantum computational tasks and resource characteristics. For example, the fundamental difference between the characteristics of a quantum task and a classical task is their information unit, i.e., quantum bits and classical bits \cite{leymann2020quantum}. Besides, current quantum computation processors can be characterized by different benchmarking metrics, such as circuit layer operations per seconds (CLOPS) and quantum volume (QV) \cite{wack2021quality}. Meanwhile, applying deep reinforcement learning (DRL) in optimizing such tasks represents a potential approach in similar resource management problems in the classical domain \cite{goudarzi2023distributed, chen2022adaptive}. It offers promising yet unexplored solutions for dynamic and efficient resource management in quantum cloud environments. 

A few studies have focused on heuristic approaches to designing quantum cloud resource management policies. Ravi et al. \cite{ravi2021quantum} analyzed quantum job characteristics of IBM Quantum cloud systems and then proposed an adaptive job scheduling approach based on a basic statistical analysis of historical data in their subsequent work \cite{ravi_adaptive_2021}. Ngoenriang et al. \cite{ngoenriang_optimal_2022} proposed a two-stage stochastic programming technique to allocate resources for distributed quantum computing. The technique aims to minimize the deployment cost and maximize quantum resource utilization while accounting for uncertainties like quantum task demands and computation power.
Kaewpuang et al. \cite{kaewpuang_stochastic_2022} proposed a new method for allocating qubits in a quantum cloud that considers the uncertainties of quantum circuit requirements and expected waiting time. The method consists of two stages: reservation and on-demand. In the reservation stage, historical data is used to determine the allocation of resources, while in the on-demand stage, actual requirements are considered. Cicconetti et al. \cite{cicconetti_resource_2022} proposed a resource allocation technique for distributed quantum computing, focusing on quantum network aspects. They used the Weighted Round Robin algorithm to assign network resources based on pre-calculated traffic flow weights. They developed a network provisioning simulator for evaluation, showing trade-offs between fairness and time complexity. 

However, these existing works do not consider the characteristics of heterogeneous quantum computing systems and circuit-based metrics of quantum tasks when designing the scheduling algorithm \cite{nguyen2024quantum}. Furthermore, none of the existing works leverage machine learning-based approaches for quantum task placement problems in cloud-based environments. Our paper fills this important gap  and contributes to the foundational knowledge and advancement of task placement strategies in the quantum cloud computing domain.

\section{System Model and Problem Formulation}
\subsection{System Model}
\label{system-model}
Figure \ref{fig_overview} represents an overview of our system model in quantum cloud computing environments, which is derived from our studies on the QSimPy \cite{nguyen2024qsimpy} toolkit for quantum cloud resource management. Since quantum applications cannot be permanently installed in a quantum computer, classical cloud resources are required to host these applications. Additionally, these applications can be made available as a service \cite{nguyen2024qfaas}. Users send task requests from their local devices to the classical cloud layer, where the quantum application is deployed. A corresponding quantum task (QTask), which comprises single or multiple quantum circuits, will be created for each incoming request. The broker (or scheduler) then makes a placement decision for each QTask based on its requirement and the current state of available quantum cloud computation resources. 

\textbf{Quantum Nodes:} The set of available quantum computation nodes (QNodes) at a quantum data center is defined as $ \mathcal{Q} = {q_1, q_2, ..., q_m}$, where $m = |\mathcal{Q}|$ indicates the number of available QNodes. We assume that each QNode has a single quantum processing unit (QPU), reflecting the current state of available quantum computers. 
Each QNode has different properties, such as qubit number ($q^w$), quantum volume (QV) ($q^v$) \cite{cross2019validating}, circuit layer operation per second (CLOPS) ($q^s$) \cite{wack2021quality}, supported gates ($q^g$), and qubit topology ($q^t$). 

\textbf{Quantum Tasks:} We consider each quantum task (QTask) to comprise a gate-based quantum circuit. Thus, a set of incoming quantum tasks can be defined as $ \Theta = \{\theta_1, \theta_2, ..., \theta_n\}$, where $n = |\Theta|$ is the number of QTasks. Each QTask $\theta_i$ has various properties, such as qubit number ($\theta^w$), circuit depth ($\theta^d$), used quantum gates ($\theta^g$), number of shots ($\theta^s$), qubit topology ($\theta^t$), and arrival time ($\theta^a$). The circuit depth ($\theta^d$) of a quantum circuit is defined as depth-1 circuit layer, where the depth of the following components can be considered as 1: a) a single-qubit gate from native gate set, b) a measurement, c) a reset, d) a 2-qubit gate from native gate set \cite{wack2021quality}.

\subsection{Problem Formulation}
The placement configuration of QTask $\theta_i \in \Theta$ can be define as $\xi_i = \{\theta_i, q_j\}$ where $q_j \in \mathcal{Q}$ and $1 \le j \le |\mathcal{Q}|$ is the selected quantum node index. If the placement fails due to a violation of resource constraints, such as placing a task to a QNode that does not have enough qubits, the broker needs to do the replacement (or rescheduling) to find another suitable QNode for the task execution.

\textbf{Completion Time Model:} 
The total completion time (or the makespan) represents the total waiting time for each request from the submission to the completion, including executing time and queueing time as follows:
\begin{equation}
    t_{\theta_i} = t^{wait}_{\theta_i} + t^{exec}_{\theta_i}
\end{equation}
where $t^{wait}_{\theta_i}$ is the queuing time (from the arrival time till the execution start time) and $t^{exec}_{\theta_i}$ is the quantum execution time. The execution time of a quantum task depends on the corresponding computer's required quantum circuit layer and QPU speed (CLOPS). Based on IBM Quantum study \cite{wack2021quality}, we use depth-1 circuit layer operation per second (D1CPS) instead of CLOPS for $q^s$, which is used to measure the number of depth-1 circuit layers (or circuit depth) of a quantum circuit that can be executed per second. The execution time of a QTask $\theta_i$ in QNode $q_j$ can be estimated as follows:
\begin{equation}
\label{eq_exec}
    t^{exec}_{\theta_i} = \dfrac{\theta_i^d\times \theta_i^s}{q_j^s}
\end{equation}
where $\theta_i^d $ is the circuit depth (or number of depth-1 circuit layers), $\theta_i^s$ is the number of shots to be executed, and  $q_j^s$ is D1CPS of the QNode. 

\textbf{Problem Statement:} Given a data center with heterogeneous quantum computation nodes (QNodes) and a continuous incoming workload (QTasks), design the task placement policy to select the most appropriate QNode for each incoming QTask object to minimize the total response time of all QTasks and mitigate the replacement frequency due to violation of the execution constraints. The objective can be defined as:
\begin{equation}
    \Omega(\Xi) = \min \ \sum_{i=1}^nt_{\theta_i}
\end{equation}
\textit{s.t.}
\begin{equation}
    C1: Size(\theta_i) = 1, \forall \theta_i \in \Theta
\end{equation}
\begin{equation}
    C2: t^{start}_{\theta_j} \ge t^{start}_{\theta_h} + t_{\theta_h} , \forall \theta_h, \theta_j \in \Theta, h < j \le |\Theta|
\end{equation}
\begin{equation}
    C3: \theta_i^w \le q_j^w , \forall \theta_i \in \Theta, q_j \in \mathcal{Q}
\end{equation}

where $C1$ specifies that each QTask can only be assigned to one QNode at the time for the execution; $C2$ indicates that the QTask $\theta_j$ can only be executed in the selected QNode after the completion of its predecessor at the same QNode (QTask $\theta_h$); and $C3$ requires the selected QNode $q_j$ need to have enough qubit for the execution of QTask $\theta_i$. If the QNode does not have enough qubits, the placement will fail and rescheduling will be necessary.

\subsection{Deep Reinforcement Learning Model}
Deep Reinforcement Learning (DRL) employs deep neural networks to tackle decision-making with high-dimensional states, formulated as Markov Decision Processes (MDP) \cite{mnih2015human}. An MDP is denoted as \( (\mathbb{S}, \mathbb{A}, \mathbb{P}, \mathbb{R}, \gamma) \), with \( \mathbb{S} \) and \( \mathbb{A} \) representing the sets of states and actions, respectively. \( \mathbb{P} \) defines the transition probabilities, \( \mathbb{R} \) is the reward function, and \( \gamma \in [0,1] \) is the discount factor, indicating the preference for future rewards. During discrete time steps \( t \), a DRL agent observes a state \( s_t \), selects an action \( a_t \) from policy \( \pi(a_t|s_t) \), and transitions to a new state \( s_{t+1} \), receiving a reward \( r_t \). The agent aims to maximize the expected return \( \mathbb{V}^\pi(s_t) = \mathbb{E}_\pi[\sum_{t} \gamma^t r_t], t \in \mathbb{T} \), the sum of discounted rewards obtained by following policy \( \pi \) from \( s_t \). The policy, often a neural network, is refined through training to optimize performance.

\textbf{State Space $\mathbb{S}$:}
A state of the agent's observation from the quantum cloud computing environment, which includes 1) information on all available QNodes and 2) information on current QTasks to be placed (or scheduled).

The feature vector of $m$ quantum nodes in $\mathcal{Q}$, each quantum node has $o$ features, at time step $t$ can be presented as:
\begin{equation}
    \mathcal{F}^{\mathcal{Q}}_t = \{f^{q_i^z} | \forall q_i \in \mathcal{Q}, 1 \le i \le m, 1 \le z \le o \}
\end{equation}
where $i$ is the index of a quantum node, and $z$ is the index of a quantum node’s feature.

The feature vector of the current quantum task $\theta_j$ in $\Theta$, with $p$ features at time step $t$ can be presented as:
\begin{equation}
    \mathcal{F}^{\theta_j}_t = \{f^{\theta_j^k} |  \theta_j \in \Theta, 1 \le k \le p \}
\end{equation}
where $k$ is the index of the current quantum task's feature.

All features of quantum nodes and quantum tasks are described in Section \ref{system-model}. Each QTask specification is only sent to the DRLQ agent as part of the state after its arrival. Therefore, the State space of the system can be defined as:
\begin{equation}
    \mathbb{S} = \{s_t | s_t = (\mathcal{F}_t^{\mathcal{Q}}, \mathcal{F}_t^{\theta_j}), \forall t \in \mathbb{T} \}
\end{equation}

\textbf{Action Space $\mathbb{A}$:} 
An action can be defined as the placement of a QTask on an available quantum node. The action $a_t$ at time step $t$ is an placement of QTask $\theta_j$ to quantum node $q_i$ can be defined as
\begin{equation}
    a_t = \xi_i = \{q_i, \theta_j\}, q_i \in \mathcal{Q}, \theta_j \in \Theta
\end{equation}
Thus, the Action space is equivalent to the set of all available quantum nodes at the data center:
\begin{equation}
    \mathbb{A} = \mathcal{Q}
\end{equation}

\textbf{Reward Function $\mathbb{R}$:} The main goal is to minimize the total completion of all incoming tasks. Besides, we also aim to mitigate task replacement attempts (or maximise the success rate of the task placement). To achieve these objectives, we define the reward $r_t$ at time step $t$ as follows:

\begin{equation}
\label{reward}
    r_t = 
\begin{cases} 
\dfrac{1}{t_{\theta_i}} \times (1 - \alpha\kappa) & \text{if } \text{done} = 1 \\
\Delta \times (1 + \alpha\kappa)  & \text{if } \text{done} = 0 
\end{cases}
\end{equation}
where $t_{\theta_i}$ is the total completion time of QTask $\theta_i$, $\alpha$ is the penalty factor, and $\kappa$ is the replacement count.
If the QTask is successful ($done =1$), the inverse value of its total completion time is assigned for the reward to encourage the policy to find a better placement that has a shorter total completion time to get a higher reward $r_t$. Otherwise, if the task execution fails for any reason, we apply a large negative value $\Delta$ for penalty and advise the policy to avoid similar action in the future. Besides, we also consider $\kappa$ - the number of replacements (or rescheduling attempts) of QTask $\theta_i$ and define a penalty factor $\alpha$ when assigning the reward in order to mitigate the replacement. The penalty factor acts as an additional discount factor when a QTask needs more than one placement to be successful and also magnifies the penalty if that QTask fails multiple times. Thus, our reward function can be used to achieve the main objective of minimizing the total completion time and reducing the number of task replacements.

\subsection{DRLQ Framework}
\label{sec:drlq}
Our DRLQ framework employs an enhanced deep reinforcement learning technique, combining Deep Q-Networks (DQN) and the Rainbow approach \cite{hessel2018rainbow} to optimize task placement in quantum cloud computing environments. 

The Deep Q-Network (DQN) algorithm, introduced by Mnih et al. \cite{mnih2013playing, mnih2015human}, represents a significant advancement in reinforcement learning, utilizing deep neural networks to approximate the action-value function $Q(s, a; \theta)$. The objective is to minimize the loss function:

\begin{equation}
L(\theta) = \mathbb{E}\left[\left(r + \gamma \max_{a'} Q(s', a'; \theta^-) - Q(s, a; \theta)\right)^2\right]
\end{equation}

where $y_i = r + \gamma \max_{a'} Q(s', a'; \theta^-)$ defines the target for a state-action pair $(s, a)$, with $r$ as the immediate reward, $\gamma$ as the discount factor, $s'$ as the subsequent state, and $\theta^-$ denotes the parameters of a target network that is periodically updated to stabilize the learning process. The overall process of the DRLQ framework can be represented in the Algorithm \ref{alg:enhanced_dqn_rllib}.

\begin{algorithm}[htbp]
\SetAlgoLined
\caption{DRLQ Framework for QTask Placement }\label{alg:enhanced_dqn_rllib}
Initialize the quantum cloud environment in QSimPy\;
Loading dataset for the QTask generator module\;
Register the environment with Ray\;
Initialize replay buffer $\mathcal{D}$ to capacity $N$ and priority replay configuration\;
Initialize action-value policy $Q$ with random weights\;
\For{each hyperparameter configuration}{
    Set hyperparameters using Ray Tune\;
    \For{episode $=1, M$}{
        Perform DQN process, defined in \cite{mnih2013playing}\;
        Adjust rewards and transitions for $n$-step\;
        Add parameter noise to network weights for exploration\;
        Update $Q$ using a distributional approach\;
    }
}
Select the best configuration from Ray Tune results\;
\end{algorithm}

First, we initialize the quantum cloud environment following the reinforcement learning setting. Due to the limitation of managing the practical quantum cloud environment setup, we utilize a simulated quantum cloud environment by utilizing the QSimPy simulator \cite{nguyen2024qsimpy} and a quantum application dataset, such as MQTBench \cite{quetschlich2023mqt}, for generating the synthetic QTask data for the environment. Then, we register the environment with Ray \cite{moritz2018ray}, a comprehensive machine-learning framework for managing the training and tuning. We utilize a replay buffer to enhance learning efficiency by decoupling consecutive training samples, thereby providing a diverse set of experiences for more robust neural network training. The replay buffer configuration and other training hyperparameters need to be defined for the hyperparameter tuning process. We leverage the Rainbow DQN approaches \cite{hessel2018rainbow} to combine all the advantages of the different DQN approaches, such as Multi-step Learning \cite{asis2018multi}, Distributional RL \cite{bellemare2017distributional}, Prioritized Replay \cite{schaul2016prioritized}, and Noisy Nets \cite{fortunato2019noisy}. These enhancements of DQN can collectively improve the training efficiency and effectiveness in quantum task placement, offering a potential approach to learning in the complex and dynamic environments of quantum cloud computing. Finally, we determine the best hyperparameter configuration based on the tuning process using Ray Tune \cite{liaw2018tune}.

\section{Performance Evaluation}
\subsection{Environment Setup}
To evaluate the performance of our DRLQ technique, we set up a simulated environment that reflects the actual quantum cloud environment due to current limitations on accessing and managing a quantum data center. We use QSimPy \cite{nguyen2024qsimpy}, a learning-centric framework derived from the iQuantum toolkit \cite{nguyen2024iquantum}, to create the quantum cloud environment. All experiments are conducted on a computation instance with 16 vCPUs and 64GB of RAM at the Melbourne Research Cloud.

We model a quantum data center with 10 heterogeneous quantum nodes, ranging from 16 qubits to 127 qubits, using quantum benchmarking metrics \cite{wack2021quality} from IBM Quantum and backend instances in Qiskit. The modeled quantum nodes included \textit{ibm\_sherbrooke, ibm\_washington, ibm\_brisbane, ibm\_osaka, ibm\_nazca, ibm\_kyoto, ibm\_cusco, ibm\_kolkata, ibm\_hanoi, ibm\_guadalupe}. We used Qiskit \cite{contributors2023qiskit} for the transpilation of circuits in incoming QTasks to the selected QNode and extracted the circuit metrics after transpilation to mimic the process when a quantum circuit reaches the quantum node for further execution.
To simulate stochastic incoming quantum tasks with metrics from actual quantum applications, we selected 12 quantum applications from the MQTBench dataset \cite{quetschlich2023mqt}, which contains quantum circuits in QASM files ranging from 2 to 50 qubits each. The selected quantum applications from the MQTBench dataset include:
\begin{enumerate}
\item Amplitude Estimation (AE)
\item Deutsch-Jozsa algorithm
\item Greenberger–Horne–Zeilinger (GHZ) state
\item Quantum Fourier Transformation
\item Entangled Quantum Fourier Transformation
\item Quantum Neural Network (QNN)
\item Quantum Phase Estimation (QPE) exact
\item Quantum Phase Estimation (QPE) inexact
\item Random circuits
\item Real Amplitudes ansatz with Random Parameters
\item Efficient SU2 ansatz with Random Parameters
\item Two Local ansatz with random parameters
\end{enumerate}
The number of shots is set to 1024 by default. We randomly select QTasks for each episode in the reinforcement learning process from the synthetic QTask dataset. The QTask arrival times were generated following a Poisson distribution. 

After setting up the Gymnasium-based environment for the quantum cloud, we used Ray RLlib \cite{liang2018rl}, an industry-grade reinforcement learning framework, to implement the proposed method of DRLQ and several baseline algorithms for performance comparison. We evaluated the performance of DRLQ against other popular heuristic approaches, including:
\begin{itemize}
    \item \textit{Greedy}: QTasks are greedily assigned to the QNode with the shortest waiting time, similar to an approach in \cite{nguyen2024qfaas}. If these tasks fail, they are assigned to the most powerful QNode (i.e., the one with the largest number of qubits).
    \item \textit{Round Robin}: QTasks are assigned to QNodes in a cyclic order, ensuring a balanced distribution of tasks across all available QNodes.
\item \textit{Random}: QTasks are randomly assigned to QNodes.
\end{itemize}

\subsection{Evaluation and Discussion}
\begin{figure}[htbp]
\centering
\includegraphics[width=3.3in]{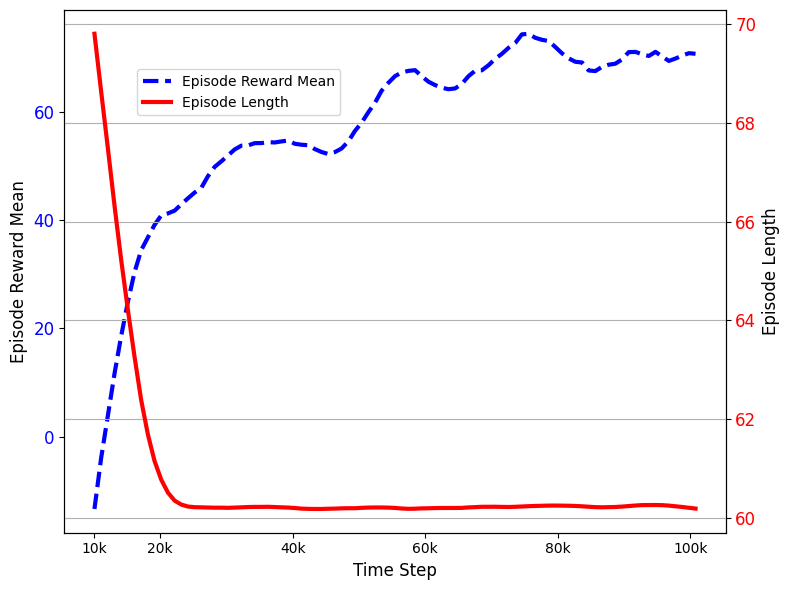}
\caption{Episode reward means and episode lengths during the training of the DRLQ policy over 100,000 time steps, total training time is 8.34 hours. Each episode consists of 60 random QTasks that arrive randomly within a 1-minute time window. The tuned hyperparameters are set as follows: learning rate (lr) = 0.01, number of atoms = 10, train batch size = 180, n\_step = 3, v\_min = -10, v\_max = 10, penalty ($\Delta$) = -10, and penalty factor ($\alpha$) = 0.1. }
\label{fig_lenreward}
\end{figure}

We used a similar evaluation approach following other DRL-based task scheduling works in classical computing, such as \cite{fan2022dras, goudarzi2023distributed}, to evaluate the performance of our framework using reward values after 100 training iterations, which involves 100,000 time steps. We conducted extensive experiments and used Ray Tune \cite{liaw2018tune} to optimize hyperparameters through the grid search method. The results of the best configuration of DRLQ are shown in Figure \ref{fig_lenreward}. 

\begin{figure*}[htbp]
  \centering
  \begin{subfigure}[b]{0.62\textwidth}
    \centering
    \includegraphics[width=\textwidth]{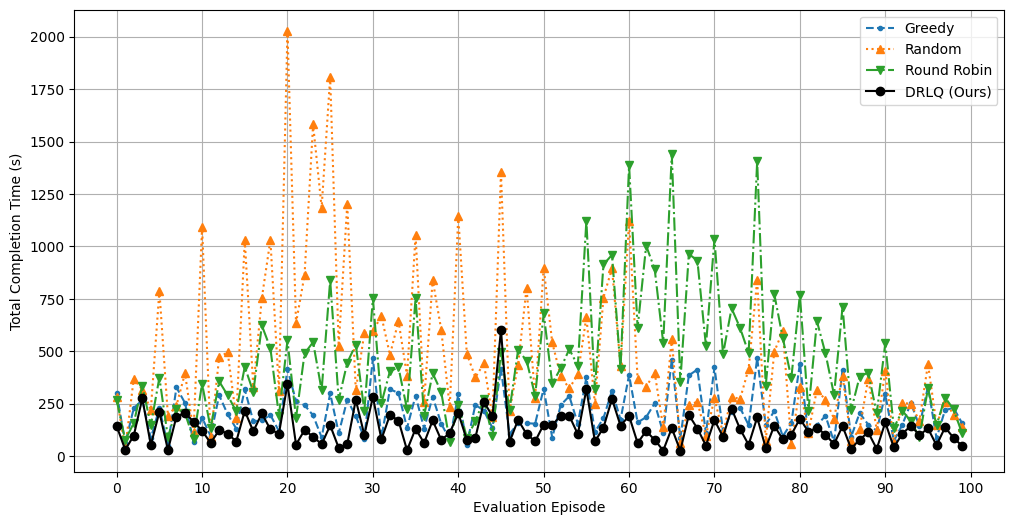}
    \caption{Total completion time of all QTasks in each episode during the evaluation.}
    \label{fig_completion}
  \end{subfigure}
  \hfill
  \begin{subfigure}[b]{0.35\textwidth}
    \centering
    \includegraphics[width=\textwidth]{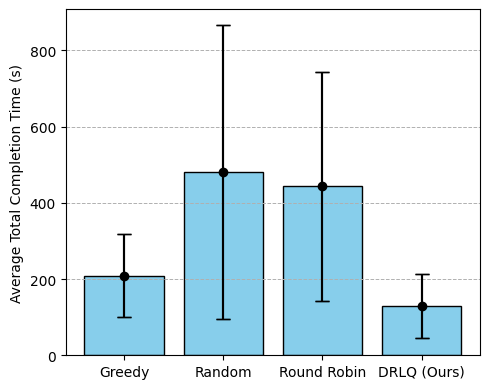}
    \caption{Average values of total completion time of all QTasks over 100 evaluation episodes.}
    \label{fig_avgcomp}
  \end{subfigure}
  \caption{Total completion times of all QTask over 100 evaluation episodes among DRLQ and other heuristic approaches.}
  \label{fig:completion_comparison}
\end{figure*}

Our main objective is to minimize the total completion time and the number of task rescheduling (or replacement) attempts. Figure \ref{fig_lenreward} clearly demonstrates the efficient learning process of the DRLQ method. As the reward is inversely proportional to the total completion time, the upward trend in the reward indicates a reduction in total completion time during the training episodes. The reward continuously increases after the first 20 training iterations, reaching convergence after 90 training iterations (each iteration consists of 1,000 time steps). Simultaneously, the episode length significantly reduces and converges around 60, which is the minimum length of an episode, after the first 25 training iterations. We then exported the trained policy after 100 training iterations for evaluation on different QTask workload datasets to compare the effectiveness of DRLQ against other heuristic approaches. 

Figure \ref{fig_completion} compares DRLQ with other baseline techniques for QTask placements, considering the total completion time of all QTasks over 100 episodes, each consisting of 60 random incoming QTasks different from the training set of DRLQ. The average total completion times of all QTasks in each episode after 100 evaluation episodes across DRLQ and other baselines are shown in Figure \ref{fig_avgcomp}. Our evaluation results indicate that the DRLQ algorithm significantly improves efficiency by minimizing the total completion time. It achieves a 37.81\% reduction in the total completion time compared to the Greedy algorithm, a 72.93\% reduction compared to the Random approach, and a 70.71\% reduction compared to the Round Robin algorithm over 100 evaluation episodes. 

\begin{figure}[htbp]
\centering
\includegraphics[width=2.6in]{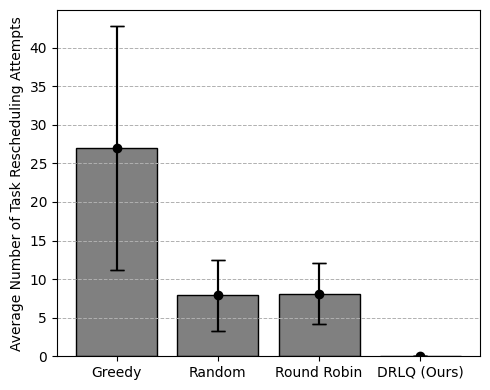}
\caption{Average number of task rescheduling attempts after 100 evaluation episodes of DRLQ and other approaches}
\label{fig_resched}
\end{figure}

We also evaluate the performance of DRLQ by considering the number of task rescheduling attempts per episode. Figure \ref{fig_resched} shows the average number of task rescheduling attempts over 100 evaluation episodes for all approaches.

The results show that our DRLQ approach significantly outperforms other methods in mitigating task rescheduling attempts. Specifically, DRLQ achieves zero rescheduling attempts, which is a substantial improvement over the Greedy, Random, and Round Robin approaches, with average rescheduling attempts of 26.95, 7.88, and 8.11, respectively. These improvements are especially important in quantum cloud computing environments, where minimizing task completion time is crucial due to the high costs of quantum resources and the inherent variability of quantum computations.


\section{Conclusions and Future Work}
In this study, we have explored the effectiveness and potential of utilizing a deep reinforcement learning approach in proposing a novel DRLQ framework for task placement in quantum cloud computing environments. Our results showcase the significant improvement of the DRLQ technique in quantum task placement compared to other heuristic approaches. It highlights the potential of using a DRL-based approach as a robust quantum cloud resource management. Our work is one of the first studies on the resource management problem of quantum cloud computing, which requests more attention from the research community in the quantum cloud domain.

Our future research directions will focus on several aspects to improve the potential of this approach. We are leveraging more advanced deep reinforcement learning techniques to improve performance in diverse quantum computing contexts. Besides, we aim to evaluate DRLQ in real quantum cloud systems to verify its practicality and fine-tune its application. We are also considering other properties of NISQ devices, such as error rates and quantum coherence, to explore the potential impacts of quantum mechanics on computational outcomes.

\section*{Acknowledgments}
The research is partially supported by the University of Melbourne through an ARC Discovery Project (awarded to Prof. Buyya). 
Hoa Nguyen acknowledges the support from the Science and Technology Scholarship Program for Overseas Study for Master’s and Doctoral Degrees, Vingroup, Vietnam. 

\bibliographystyle{ieeetr}
\bibliography{bibliography}

\end{document}